\documentstyle[pre,aps,multicol,epsf]{revtex}
\def\B.#1{{\bbox{#1}}}
\begin{document}
\title{
Outliers, Extreme   Events and Multiscaling} 
\author{Victor S. L'vov, Anna Pomyalov and Itamar Procaccia} 
\address{Dept. of Chemical Physics, The Weizmann
Institute of Science, Rehovot 76100, Israel}   
\maketitle

\begin{abstract}
Extreme events have an important role  which is sometime catastrophic
in a variety of natural phenomena including climate, earthquakes and
turbulence, as well as in man-made environments like financial
markets. Statistical analysis and predictions in such systems are
complicated by the fact that on the one hand extreme events may appear
as ``outliers" whose statistical properties do not seem to conform
with the bulk of the data, and on the other hands they dominate the
(fat) tails of probability distributions and the scaling of high
moments, leading to ``abnormal" or ``multi"-scaling.  We employ a
shell model of turbulence to show that it is very useful to examine in
detail the dynamics of onset and demise of extreme events.  Doing so
may reveal dynamical scaling properties of the extreme events that are
characteristic to them, and not shared by the bulk of the
fluctuations. As the extreme events dominate the tails of the
distribution functions, knowledge of their dynamical scaling
properties can be turned into a prediction of the functional form of
the tails. We show that from the analysis of relatively short time
horizons (in which the extreme events appear as outliers) we can
predict the tails of the probability distribution functions, in
agreement with data collected in very much longer time horizons. The
conclusion is that events that may appear unpredictable on relatively
short time horizons are actually a consistent part of a multiscaling
statistics on longer time horizons.
\end{abstract}
\begin{multicols}{2}
\section{Introduction}
\label{s:intro}
There is an obvious and wide spread interest in predicting extreme
events in a variety of contexts. Particularly well known examples are
the insurance risks related to large tropical storms, human and
property risks in the context of large earthquakes, financial risks
caused by large movements of the markets, and dangers to passenger
planes due to extremely intermittent turbulent air
velocities. Obviously, any improvement in the predictability of any of
these extreme events is highly desirable for a number of
reasons. Accordingly, there exists a large body of work focusing on
the statistics of such events, small, intermediate and large, with the
aim of studying the ensuing probability distribution functions
(PDF). If one can model properly the PDF, one can in principle predict
at least the frequency of extreme events. Yet, there is one
fundamental question that arises that needs to be confronted first:
are the extreme events sharing the same statistical properties as the
small and intermediate events, or are they "outliers"? If the latter
is true, then no analysis of the core of the PDF, clever as it may be,
could yield a proper answer to the desire to predict the probability
of extreme events.

Indeed, in a number of context it had been proposed recently that
extreme events are ``outliers" \cite{99Sor}. For example in financial
markets the largest draw-downs appear to exhibit properties that
differ from the bulk of the fluctuations \cite{DowJones}. In general
one would refer to ``outliers" when the rate of occurrence of small
and intermediate events lies on a PDF with some given properties,
while the extreme events appear to exhibit statistical properties that
differ from the bulk in a significant way.  The aim of this paper is
to present a detailed analysis of the fluctuations in a turbulent
dynamical system that shows that such a point of view can be
substantiated . Clearly, this type of considerations must be conducted
with great care. The danger is that on small time horizons the largest
events appear so rarely, once or twice, that their rate of occurrence
is not statistically significant, and no conclusion about their
relation to the statistics of small and intermediate events is
possible. Nevertheless, we offer in this paper a positive outlook. We
will show that in the context of the bulk of this paper, which is the
analysis of a shell model of turbulence, one can analyze {\em within
the short time horizon} the dynamics of the extreme events. This
analysis reveals their special dynamical scaling properties, allowing
us to make interesting predictions about the tails of the distribution
functions even before the full statistics is available.  These
predictions can be checked in our case by considering much longer time
horizons. The conclusion for the extreme events community is that it
may very well pay to look very carefully at the detailed dynamics of
the extreme events if one wants to claim anything about their
probability of occurrence.

The model that we treat in detail in this paper is a so-called
``shell" model of turbulence. Shell models of turbulence
\cite{Gledzer,GOY,Jensen91PRA,Piss93PFA,Benzi93PHD,sabra} are
simplified caricatures of the equations of fluid mechanics in
wave-vector representation; typically they exhibit anomalous scaling
even though their nonlinear interactions are local in wavenumber
space. The wavenumbers are represented as shells, which are chosen as
a geometric progression
 \begin{equation}\label{kn} k_n=k_0 \lambda^n\,, 
\end{equation} 
where $\lambda$ is the ``shell spacing". There are $N$  degrees of
freedom where $N$ is the number of shells. The model specifies the
dynamics of the ``velocity" $u_n$ which is considered a complex
number, $n=1,\dots,N$.  Their main advantage is that they can be
studied via fast and accurate numerical simulations, in which the
values of the scaling exponents can be determined very precisely. We
employ our own home-made shell model which had been christened the Sabra
model\cite{sabra}. It exhibits similar anomalies of the scaling
exponents to those found in the previously popular GOY model
\cite{Gledzer,GOY}, but with much simpler correlation properties, and
much better scaling behavior in the inertial range.  The equations of
motion for the Sabra model read:
\begin{eqnarray} \label{sabra}
\frac{d u_n}{dt}&=&i\big( ak_{n+1}  u_{n+2}u_{n+1}^*
 + bk_n u_{n+1}u_{n-1}^*  \\ \nonumber
&& -ck_{n-1} u_{n-1}u_{n-2}\big)  -\nu k_n^2  u_n +f_n\,,
\end{eqnarray}
where the star stands for complex conjugation, $f_n$ is a forcing term
which is restricted to the first shells and $\nu$ is the ``viscosity".
In this paper we restrict the forcing
to the first and and second shells only ($n=1,2)$.  
The coefficients $a,b $ and $c$ are chosen such that
\begin{equation}\label{abc}
a+b+c=0 \ .
\end{equation}
This sum rule guarantees the conservation of the ``energy"
\begin{equation}\label{energy}
E=\sum_n |u_n|^2\,, 
\end{equation} 
in the inviscid ($\nu=0$) limit.

The main attraction of this model is that it displays multiscaling in
the sense that moments of the velocity depend on $k_n$ as power laws
with nontrivial exponents:
\begin{equation}
\label{scaling}
S_p(k_n)\equiv \langle |u_n|^p\rangle \propto k_n^{-\zeta_p} \propto
\lambda^{-n \zeta_p}\,,
\end{equation}
 where the scaling exponents $ \zeta_p$  exhibit non linear dependence
 on $p$.  We expect such scaling laws to appear in the ``inertial
 range" with shell index $n$ larger than the largest shell index that
 is effected by the forcing, denoted as $n_L$, and smaller than the
 shell indices affected by the viscosity, the smallest of which will
 be denoted as $n_d$. The scaling exponents were determined with high
 numerical accuracy better than  $0.02$ in Ref.\cite{sabra}.

\begin{figure}
\epsfxsize=9cm
\epsfbox{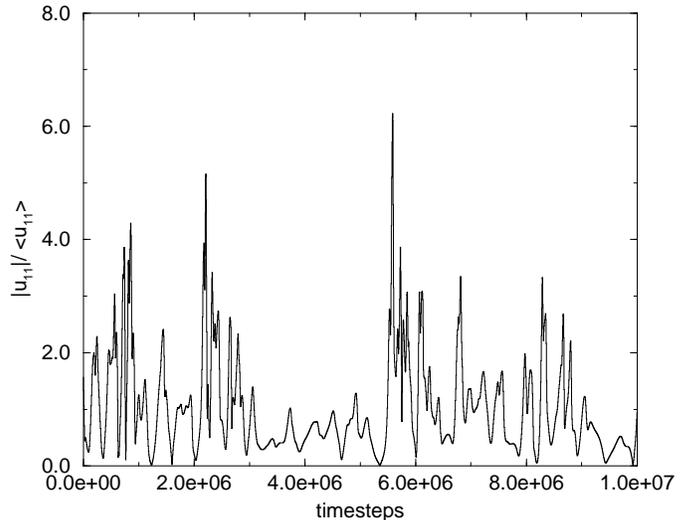}
\caption{Time series for normalized  velocity of the 11th
  shell. Parameters of the numerics: $a=1$, $b=c=-0.5$, $\lambda=2$,
  $N=28$, $k_0=1/64$, time correlated random forcing on the first two
  shells with characteristic amplitude $0.005(1+i)$. Decorrelation
  time is chosen about turnover time of the 1st shell. }
\label{fig1}
\end{figure}

\begin{figure}
\epsfxsize=9cm
\epsfbox{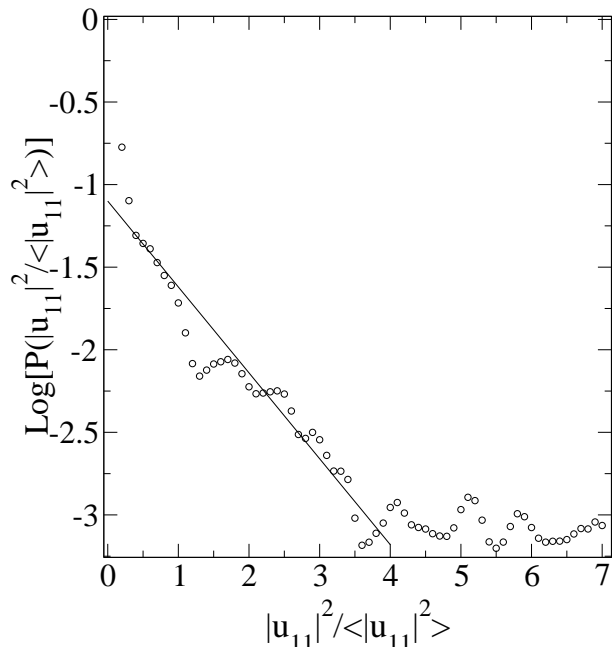}
\caption{Apparent probability distribution function for 11th shell.
Averaging over $10^7$ time-steps which is about 250 decorrelation
times for this shell. The data contains additional, extremely
sparse events of amplitude larger than 7, occurring once each; these
were left out in this plot.}
\label{fig2}
\end{figure}

To introduce the issue behind the title of this paper, we present in
Fig.~\ref{fig1} a typical time series for $u_{11}$.  The parameters of
the model are detailed in the figure legend. One can see the typical
appearance of rare events with amplitude that exceeds the mean by a
factor of 6--8.  To pose the question in its clearest way we display
in Fig.~\ref{fig2} a distribution function which is the normalized
rate of occurrence (i.e. the number of times) that a given amplitude
has been observed in the time window of $10^7$ time steps. This
apparent relative frequency of events is very similar to findings in
real data, see for example Fig.~1 of Ref.~\cite{DowJones}. which deals
with draw downs in the Dow Jones Average. Similarly to the analysis
there, we can pass an approximate straight line through the points
representing small and intermediate events. Such an exponential law
would mean that the events of $|u_{11}|^2$ with amplitudes larger
than, say, 4$\langle|u_{11}|^2\rangle$ are clear outliers. Their
probability is so low that they should not have appeared
in the short time horizon at all. We could conclude, like in the
 analysis of Ref.~\cite{DowJones}, that the extreme events cannot be
 dealt with the same distribution function as the small and
 intermediate events.
~\vskip -1.2cm

On the other hand, it is very possible that the low rate of occurrence
of the extreme events in Fig.~\ref{fig2} means simply that they are
statistically irrelevant and that no conclusion can be drawn. How to
overcome this difficulty? The purpose of this paper is to show that
indeed the extreme events may have dynamical scaling properties that
are all their own, and that they affect crucially the tails of the
distributions functions, making them very broad indeed. The main new
point is that detailed analysis of the extreme events {\em in the
short time horizon} suffices to make lots of predictions about the
tails of the PDF's, predictions that in our case can be easily
confirmed by considering much longer time horizons.

In explaining our ideas we will try to distinguish aspects which are
general, and that in our view may have applications to other systems
with extreme events, and aspects which are particular to the example
of the shell model of turbulence. Thus we start in Sect.2 with an
analysis of the temporal shape of the extreme events. We believe that
this analysis is very general, leading to an important relation
between the amplitude of the event and its time scale (the time
elapsing from rise up to demise).  In Sect. 3 we employ the
dynamical scaling form of the extreme events to present a theory
of the tails of the distribution functions. We can relate the tails
of PDF's belonging to different scales. In Sect. 4 we discuss numerical
studies of the PDF's, distinguishing the core and the tails. In Sect.5
the main numerical findings are rationalized theoretically on the
basis of universal ``pulse" solutions of the dynamics of the
Sabra model.
Section 6 contains the bottom line: we make use
of the scaling relations to {\em predict} the tails of PDF's from
data collected within short time horizons. Direct measurements
of these tails give nonsense unless the time horizons are 
increased a hundred fold. Yet with the help of the theoretical
forms we can offer predicted tails that agree very well with
the data collected with much longer time horizons.

\section{Detailed dynamics and scaling of the extreme events}
\label{s:dynamics}

In turbulence in general and in our shell model in particular the
energy that is injected by the forcing at the largest scales ($n=1$
and 2) is transferred on the average to smaller scales. It is
advantageous to analyze the extreme events of a given scale (or given
shell $n$) and also to follow the cascade of extreme events from scale
to scale. We first consider a given shell.
\subsection{Temporal dependence of extreme events of a given  scale}
\label{ss:samescale}
\begin{figure}
\epsfxsize=8.3cm \epsfbox{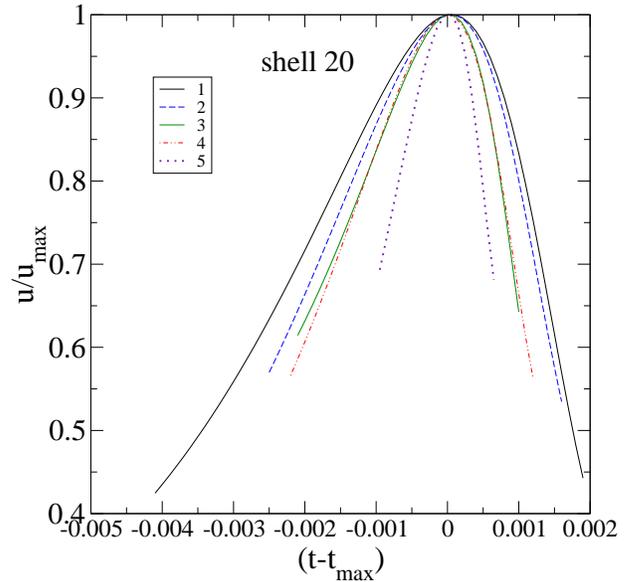}
\caption{Collapse (of positions and amplitudes) for five  intensive 
peaks  belonging to 20th shell. The values of $u_{\rm max}$ for
the peaks numbered from 1 to 5 are 4.65, 4.77, 6.71, 7.40 and
10.5 respectively, in units of the rms velocity in this shell.
The narrowest peak is thus the tallest.}
\label{fig3-1}
\end{figure}

We focus here on the detailed dynamics of the largest events of a
given scale. We considered for example the time series of the 20th
shell ($n=20$) and isolated the 5 largest events (in terms of their
amplitude) as they occurred in a time window of $10^7$ time steps. In
the first step of analysis we normalized these 5 events by the
amplitude at their maximum. Next we plotted these normalized events as
a function of time, subtracting the time at which they have reached
their maximum value. The result of this replotting is shown in
Fig.~\ref{fig3-1}. Obviously a similar replotting can be done for any
time series, and by itself is contentless.

~\vskip -1.2cm
\vskip -0.8cm
\begin{figure}
\epsfxsize=8.3cm \epsfbox{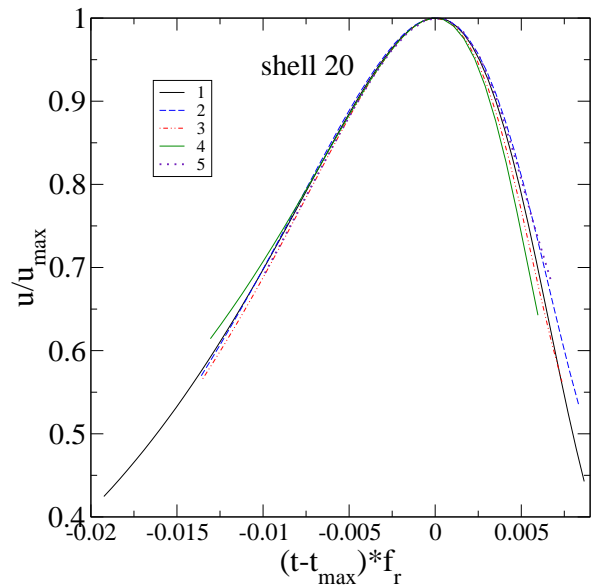}
\caption{Full collapse (of the position, amplitude and width) 
of the same as in Fig.~\ref{fig3-1} peaks. The ordering of the points
is 1 to 5 from left to right.}
\label{fig3-2}
\end{figure}

The next step of analysis will reveal already something interesting.
Building on the normalized events of Fig.~\ref{fig3-1} we attempt to
rescale the time axis for each event in order to collapse the data
together. Of course, each event calls for a different rescaling
factor, which we denote (in frequency units) as $f_r$.  The fact that
such a rescaling factors exist, and that they leads to data collapse
as shown in Fig.~\ref{fig3-2}, is a not trivial fact which may or may
not exist in different cases. But we will show that if such a
rescaling is found, it can serve as a starting point for very useful
considerations.

\begin{figure}
\epsfxsize=8.6cm 
\epsfbox{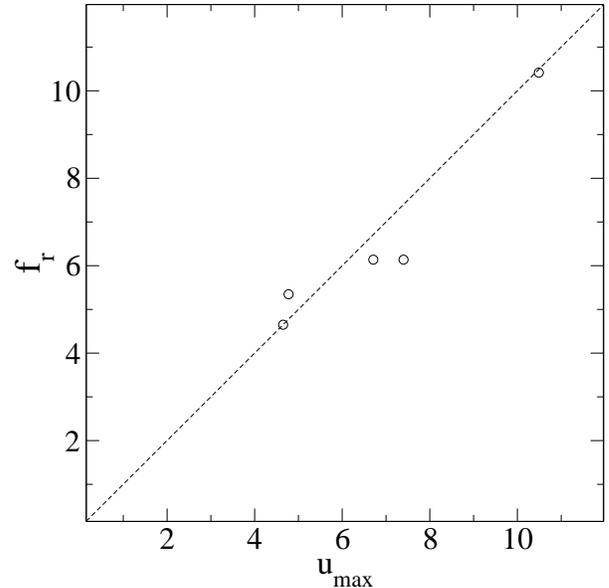}
\caption{Width normalization {\sl vs} amplitude for 5 peaks collapsed
  on Fig.~\ref{fig3-2}.}
\label{fig3-3}
\end{figure}

The third step of the present analysis is a search of meaning to the
rescaling factors $f_r$. We hope that $f_r$ has a simple relation to
the amplitude of the extreme events. To test this we can plot the
individual values of $f_r$ found in Fig.~\ref{fig3-2} as a function of
the amplitude at the peak. The resulting plot is shown as
Fig.~\ref{fig3-3}. In passing the straight line through the data
points we included the point $(0,0)$ in the analysis, as we search for
a simple scaling form
\begin{equation}
f_r \propto u_{\rm max}^x \ .
\end{equation}
with $x$ a scaling exponent. We conclude that in this case
we have a satisfactory scaling law with $x=1$.

The meaning of this scaling law is quite apparent in the present
case. Looking back at the equation of motion we realize that from the
point of view of power counting (not to be confused with
actual dynamics) it can be written as
\begin{equation}
\frac{d u}{d t} \propto u^{1+x}
\end{equation}
with $x=1$. It is thus acceptable that a rescaling of $1/t$
by $u_{\rm max}$ should collapse all the extreme events
as shown above. If the equation of motion were cubic in $u$ we
could expect $x=2$ etc. Obviously, the rescaling analysis
in this case revealed the type of dynamics underlying the
process. Whether this can be done effectively in other case
where extreme events are crucial is an open question for
future research.

\begin{figure} 
\epsfxsize=8.6cm \epsfbox{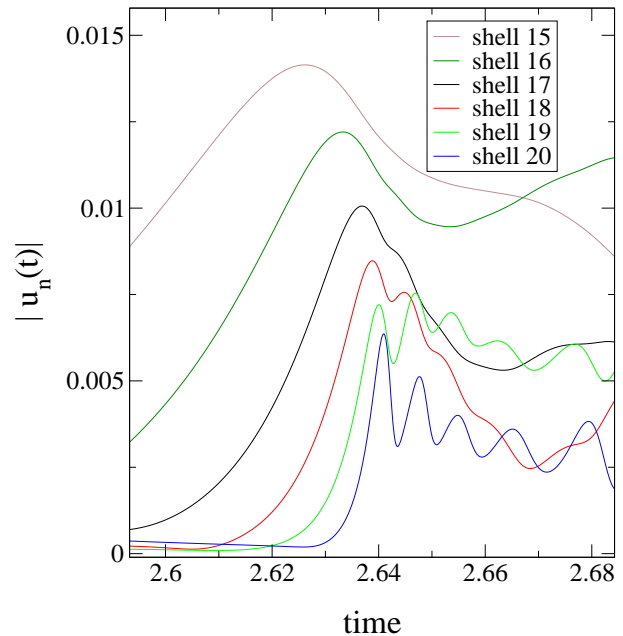}
\caption{``Evolution'' of a  
peak from the 15th to the 20th shell. The amplitudes are all in the
same (arbitrary) units.  One sees a progressive shift of the maximum
to the right and a decrease in the amplitude, accompanied by narrowing
and splitting. Nevertheless the form of the central part of the peak
remains self-similar as exempified in the two following figures.}
\label{fig4}
\end{figure}

\subsection{Transfer of extreme events between different scales}
\label{ss:diffscale}

To gain further understanding of the extreme events we focus now on
the transfer from scale to scale. Consider for example a particular
large amplitude event in the shell $n=15$, and its future fate as time
proceeds. This is shown in Fig.~\ref{fig4}. The event reached its
highest amplitude at shell 15 around $t=2.625$. At a slightly later
time it appeared as a large event in shell 16, and with a shorter
delay at shell 17 where it started to split into a doublet.  At even
shorter delays this event emerges as a triplet and a multiplet at
shells 18,19 and 20 respectively.

\begin{figure}
\epsfxsize=8.6cm \epsfbox{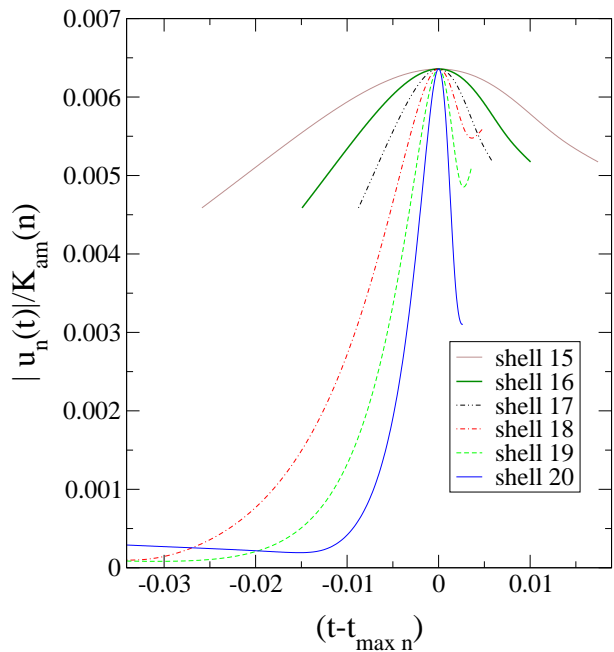}
\caption{Collapse of the  peak  amplitudes for 15--20  shells. Initial
  peaks are shown on Fig.~\ref{fig4}. }
\label{fig5}
\end{figure}

A very important characteristic of the dynamics of large events can be
obtained from finding how to relate the maximal amplitudes of the
first peak in the different shells.  As was done above, we first
replot all the first peaks as a function of time minus the time $t_n$
of their maximal amplitude $u_{n, \rm max}$.  We then glue all the
maxima together by rescaling the peaks amplitudes relative to the peak
of a chosen shell. Denote by $K_{\rm am}(n,m)$ the relative amplitude
of the peak in the $n$th shell to the $m$th shell. Choosing in our
example $m=20$ we then seek a single exponent $y$ such that
\begin{equation}
  \label{eq:rescale}
K_{\rm am}(n,20)\equiv u_{n, \rm max}/ u_{20, \rm max}=
\lambda^{(20-n)y} \,,
\end{equation}
where $\lambda$ is the shell spacing defined by Eq.~(\ref{kn}).  The
value of $y$ is obtained by plotting $g_{\rm am}(n)$ {\sl vs} $(20-n)$
where 
\begin{equation}
  \label{eq:gam}
g_{\rm am}(n)\equiv \ln [K_{\rm am}(n,20)]/\ln\lambda=y\, (20-n)\ .
\end{equation} 
The best fit is obtained with $y = 0.24 \pm 0.01$, see
Fig.~\ref{interfit}.  The peaks which are now glued at their maxima as
shown in Fig.~\ref{fig5} still have very different time-width.

\begin{figure}
\epsfxsize=8.6cm
\epsfbox{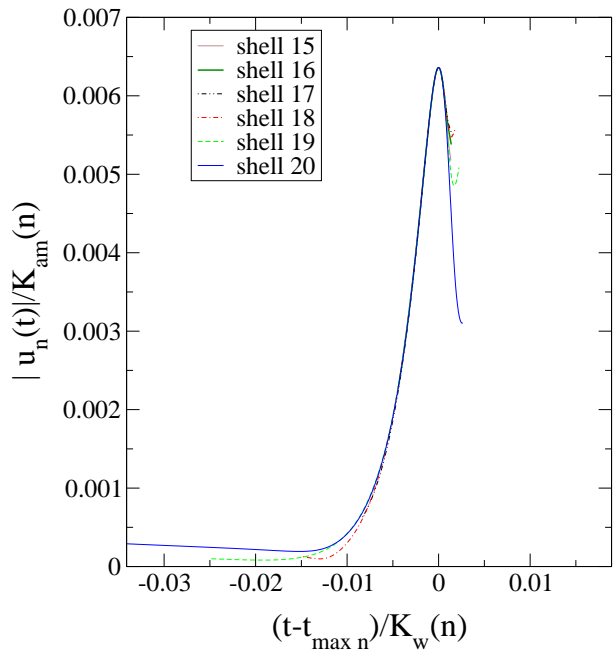}
\caption{Full self-similar collapse of  the peaks for 15--20 
shells.}
\label{fig6}
\end{figure}

\begin{figure}
\epsfxsize=8.6cm \epsfbox{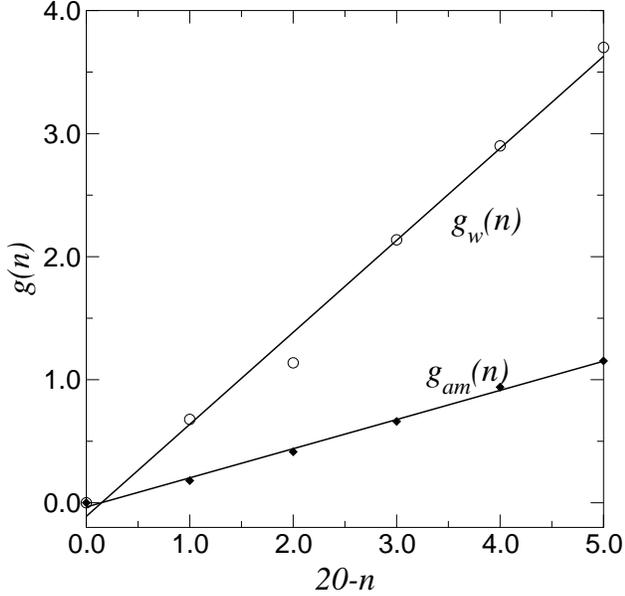}
\caption{Fits of the rescaling factors $g_{\rm w}(n)$ and  $g_{\rm
    am}(n)$ for the peaks in the shells 15 -- 20 shown in
    Figs.~\ref{fig4},\ref{fig5} and \ref{fig6}. Note that
in comparing {\em different} shells the rescaling of the frequency
increases when the peak decreases in amplitude. This is opposite
to the rescaling of peaks {\em within} a given shell.}
\label{interfit}
\end{figure}

Next, as before, we want to collapse all these curves by rescaling the
time axis according to $(t-t_n) \to (t-t_n) / K_{\rm
w}(n,20)$. Expecting the scaling law $K_{\rm
w}(n,20)=\lambda^{z(20-n)}$ it is natural to consider
\begin{equation}
  \label{eq:gw}
g_{\rm w}(n)\equiv \ln [K_{\rm w}(n,20)]/\ln\lambda=z\, (20-n)\ .
\end{equation} 
The exponent $z=0.75 \pm 0.02$ is found by computing ``the best''
linear fit of $g_{\rm w}(n)$ {\sl vs} $(20-n)$, see
Fig.~\ref{interfit}. The quality of the resulting data collapse can be
seen in Fig.~\ref{fig6}.  Note, that within the error bars
$z+y=1$. This sum rule will be rationalized theoretically in
Sec.~\ref{s:theory}.

The bottom line of this analysis can be summarized in a dynamical
scaling form for the extreme events:
\begin{equation}
u_n(t)\approx v\lambda^{-yn} f\left[(t-t_n)vk_0 
\lambda^{zn} \right] \ .
\label{scform}
\end{equation}
Here $v$ is a characteristic velocity amplitude associated with the
cascade of a particular large event which starts at small $n$ and
reaches eventually large values of $n$.  As such $v$ is not
universal. We stress that the scaling form was derived on the basis of
a time series in the short time horizon, i.e. the the same one that
gave rise to the apparent PDF shown in Fig.~\ref{fig1}. We will see
that these findings suffice to make rather strong predictions about
the expected form of the {\em converged} PDF. A theoretical
understanding of the origin of the scaling form (\ref{scform}) will be
presented in Sec.~\ref{s:theory}.
 ~\vskip -0.1cm
\vskip -1cm

\begin{figure}
\epsfxsize=8.6cm \epsfbox{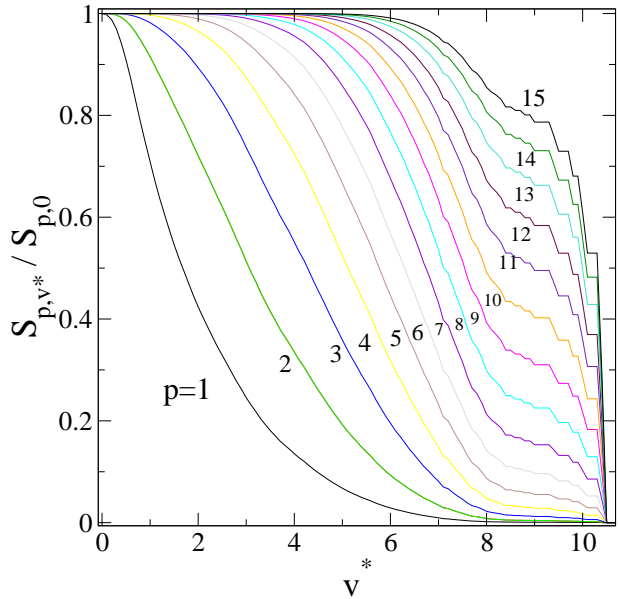}
\caption{Normalized contributions  to the structure functions of 
orders $p=1,\ 2\ \dots 15$ for 20th shell from the part of the
velocity realization with $v>v_*$ }
\label{fig10}
\end{figure}

\section{Implications for the tails of the Probability Distribution
 Functions}
\label{s:predPDF}
\subsection{Asymptotic Scaling Exponents}
\label{ss:model}

Having a scaling form for the large events means a great deal for the
structure functions $S_p(k_n)$ [cf. Eq.~(\ref{scaling})] for high
values of $p$. In fact for high $p$ the structure functions are
dominated by the large events. To demonstrate this we show in
Fig.~\ref{fig10} the relative contribution to $S_p(k_{20})$ that arise
from velocity amplitudes that exceed a threshold $v_*$. In this plot
$S_{p,v_*}$ is the structure function Eq.~(\ref{scaling}) where only
events with $u_{20}\ge v_*$ are considered, whereas $S_{p,0}$ contains
all the data. Obviously the higher $p$ is the higher is the
contribution of large events. For any time window there exists the
largest event, and when $v_*$ exceeds its value, $S_{p,v_*}$
necessarily vanishes.

If we accept the scaling form~(\ref{scform}) we can use it
to predict the scaling exponent $\zeta_p$ for high values
of $p$. By definitions
\begin{equation}
S_p(k_n) = \lim_{T\to \infty}\frac{1}{2T} \int\limits _{-T}^T
|u_n|^p dt \propto k_n^{-\zeta_p}\propto \lambda^{-n\zeta_n}\ .
\label{defSp}
\end{equation}
For $p$ large enough the structure functions are dominated by the well
separated events. Instead of the integral in the interval $[-T,T]$ we
can sum up the inegrals over the separated peaks. Substituting for
each peak the form ~(\ref{scform}) and noting that the number of peaks
is proportional to $T$, we can extend the integration interval to
$[-\infty,\infty]$ and write
\begin{eqnarray}\label{pred-Sp}
S_p(k_n) &\propto& \lambda^{-ynp}\!\!
\int\limits_{-\infty}^\infty \!\! f^p(\lambda^{zn}t v k_0) dt\\ 
\nonumber &\propto&
\lambda^{-n(yp+z)} \!\! \int \limits_{-\infty}^\infty \!\!
f^p(\tau)d\tau \ .
\end{eqnarray}
Comparing the exponents of $\lambda$ here and in the previous
equation we find the
scaling exponents
\begin{equation}\label{pred-zetap}
\zeta_p =yp+z \ .
\end{equation}
Of course this prediction is valid only for high values of $p$
for which the contributions of the isolated peaks are domninant.
\subsection{Tails of the Probability Distribution Function}

We turn now to the prediction of the tails of the PDFs assuming that
these tails are dominated by well separated peaks with self-similar
evolution~(\ref{scform}). We will see below [and cf.
Eq.~(\ref{pred1-PDF})], that the tails of the predicted PDF are very
sensitive to the {\em exponents} in Eq.~(\ref{scform}), but rather
insensitive to the precise form of the universal function $f(x) $ in
Eq.(\ref{pred-Sp}).  Assume then for simplicity that $f(x)=1$ for
$|x|\le 1/2$ and $f(x)=0$ for $|x| > 1/2$.  There is the free
parameter $v$ in Eq.~(\ref{scform}); for the chaotic realizations
$u_n(t)$ we consider it as a random parameter. Define then the
variable $V^2$ according to
\begin{equation}
  \label{eq:V2}
 V^2\equiv v^2/v_0^2\,,\quad v_0^2=\sum_{n=1}^\infty \langle
 u_n^2\rangle \ .
\end{equation}
Consider now a run with a total time horizon $T\equiv 1/ (k_0
v_0)$. Denote as $W(V^2)dV^2$ the number of peaks measured in this run
in which the value of $V^2$ fell in the window $[V^2 ,(V^2+dV^2)]$.

Next denote normalized amplitudes [the value of the signal at times
$t=t_n$ in Eq.(\ref{scform})]
\begin{equation}
  \label{eq:U2}
  U_n^2\equiv \frac{u_n^2}{ \langle u_n^2 \rangle}=\frac{V^2}{C
  \lambda^{\alpha n}}\,, \quad \alpha\equiv 2y-\zeta_2\,,
\end{equation}
where $C$ is a dimensionless constant.  We are interested in the PDF
$P_n(U_n^2)$, where $P_n(U_n^2)dU_n^2$ is the probability to sample a
normalized amplitude in the $n$th shell between $U_n^2$ and
$U_n^2+dU_n^2$. By definition, the number of observations of such
amplitudes in the time horizon $T$ is $dN_n$
\begin{equation}
dN_n = P_n(U_n^2)dU_n^2\frac{T}{\tau_0}\ , \label{dNn}
\end{equation}
where $\tau_0$ is the length of the sampling intervals. On the
other hand, since the lifetime of a peak with a given value of 
$V^2$ belonging to the $n$th shell is $1/vk_0\lambda^{zn}$, we
can also estimate the number of observations $dN_n$ as
\begin{equation}
  \label{eq:dN1}
  dN_n=\frac{W(V^2)}{\tau_0 v k_0 \lambda^{n z }}dV^2 \ .
\end{equation}
Equating Eqs.(\ref{dNn}) and (\ref{eq:dN1}) and rearranging,
one gets:
\begin{equation}
  \label{pred1-PDF}
  P_n(U_n^2)=C\, W(V^2)\lambda^{n(\alpha-z)}/V \ .
\end{equation}
This relation is obtained under the assumption that the number of
peaks is not increasing in the cascade process. In fact we saw that
the number of the peaks is increasing with the shell number $n$,
presumably in a scale-invariant manner as $\lambda$ to some positive
exponent $\beta$. We can account for this effect by replacing in
Eq.~(\ref{pred1-PDF}) $W$ by $\lambda^\beta \,W$. After that:
\begin{equation}
  \label{pred2-PDF}
  P_n(U_n^2 )=C\,W(V^2)\lambda^{n(\alpha+\beta-z)}/V \,,
\end{equation}
where $V^2$ and $U_n^2$ are defined by Eqs.~(\ref{eq:V2})
and~(\ref{eq:U2}).  Equation~(\ref{pred2-PDF}) means that a collapse
of the tails of the PDFs for different shells may be achieved by
rescaling the $x$-axis $U_n^2 \to V^2$ according to~(\ref{eq:U2}) and
rescaling of the PDFs ($y$-axis) by $\lambda^{n(\alpha+\beta-z)}$.

Equation~(\ref{pred2-PDF}) for the tail of the PDFs allows one to
find the high order structure functions (which are dominated by the
tails of the PDFs) and their scaling exponents $\zeta_p$:
\begin{eqnarray}
  \label{eq:Sp}
    S_p(k_n)&=&\int\limits_0^\infty u_n^p P_n(U_n^2)\, d\, U_n^2
    =C_p\, v_0^p\,\lambda^{n (\beta-z-yp )}\,, \\ \label{eq:tz} C_p
    &=&\int\limits_0^\infty V^{p-1} W(V^2)\,\,V^2 \ .
\end{eqnarray}
Comparing again the exponents of $\lambda$ here and in Eq.(\ref{defSp})
gives the prediction for the high order scaling exponents:
\begin{equation}
  \label{eq:zp}
  \zeta_p=yp+z-\beta\,,
\end{equation}
which coincides with Eq.~(\ref{pred-zetap}) at $\beta=0$. One sees that
the effect of peak splitting (which was  described by positive exponent
$\beta$) increases the deviation of the scaling exponents from its K41
value $\zeta_p=p/3$.
\section{Numerical studies of the PDF: core and tail}
\label{s:check}
\begin{figure}
\epsfxsize=9.0truecm \epsfbox{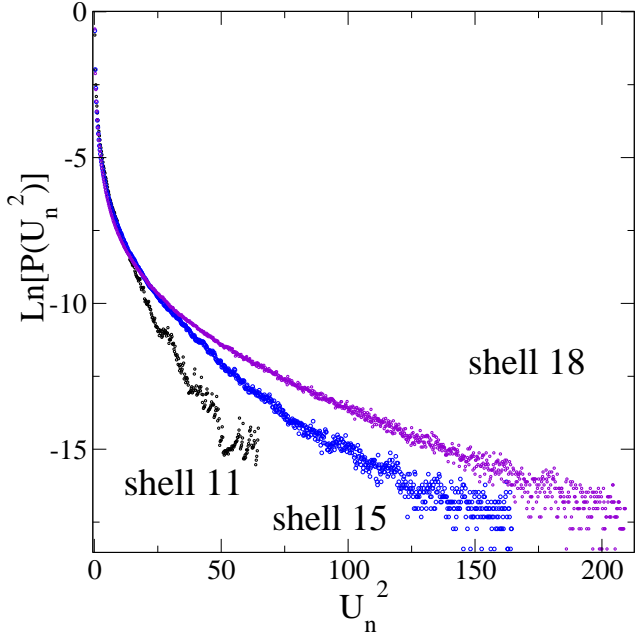}
\epsfxsize=9.0truecm \epsfbox{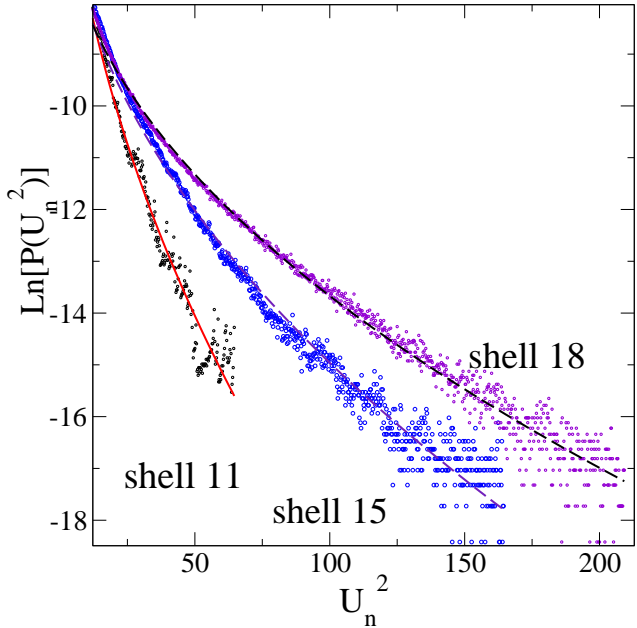}
\caption{Upper panel: PDFs of the 11th, 15th and 18th shells 
(averaged over $10^9$ time steps). Lower panel: Tails of PDFs 
(with the cores left out) fitted
by functions of the form $\ln[P_n(U_n^2)]=a_n + b_n U_n$ (
continuous lines).}
\label{fig:sh11-18}     
\end{figure}

It is well known that PDF's in multiscaling systems are not scale
invariant. Nevertheless we need to examinte the possibility that the
cores of the PDF's can be collapsed using a rescaling law that is
charateristic to them, while the tails may be collapsed using another
rescaling law (with different scaling exponents).  This possibility is
related to the fact that the structure functions $S_p(k_n)$ have
scaling exponents in the vicinity of the K41 values ($\zeta_{_{\rm
K41}}(p)=p/3$) for $p$ small enough, [say $p\le 6 $].  For large value
of $p$ (say $p>12$) the $p$-dependence of $\zeta_p$ has a different
slope, cf. Eq~(\ref{eq:zp}). These differences result from the core of
PDFs originating from the bulk of the fluctuations while the tail of
PDFs resulting from the well-separated high amplitude
peaks. Accordingly the functional form of the core and the tail of the
PDFs are different. This is demonstrated in Fig.~\ref{fig:sh11-18}
(upper panel) where the PDFs for the 11th, 15th and 18th shells are
displayed. One sees that the cores (say $U_n^2\le 20$) are practically
collapsed while the tails are widely separated. Needless to say, the
collapse is due to our choice of display as a function of $U_n^2$: for
K41 PDF's such a display would result in a complete collapse, core as
well as tail.  We stress though that if one exapnded the scale one
could observe that the collapse of the core is not precise: the
scaling exponents even for $p=2$ and $p=4$ are {\em not} 2/3 and 4/3
respectively. The anomaly of these exponents is however sufficiently
small to allow an approximate collapse of the cores.
Our aim here is to test the predictions regarding the tails of the
PDF's. We note that PDFs that originate from data tend to have rather
noisy tails. This poses difficulties in assessing the accuracy of the
collapse of the tails.  Therefore we opt to first fit the PDFs with
some appropriate functional form and then to collapse the fit
functions. As a natural fit function we choose $\ln[P_n(U_n^2)]=a_n +
b _n U_n^{2 c_n}$ with three free parameters $a_n$, $b_n$ and $c_n
$. The results of our fits showed that the parameters $c_n$ are close
to $1/2$ for all values of $n\ge 11$.  Therefore we fixed the value
$c_n=1/2$ and optimized the values of of $a_n$ and $b_n$ to get the
best fits in the tail regions. Now the fit formula reads
\begin{equation}
  \label{eq:fits}
  \ln[P_n(U_n^2)]=a _n + b _n U_n  \ .
\end{equation}
The corresponding fits for the tails of the PDFs for the 11th, 15th and
18th shells are shown in Fig.~\ref{fig:sh11-18}, lower panel.  The
fits are excellent for $U_n^2> 20$ but not surprisingly they fail for
smaller values of $U_n^2$, especially for larger value of $n$.

\begin{figure}
\epsfxsize=9truecm 
\epsfbox{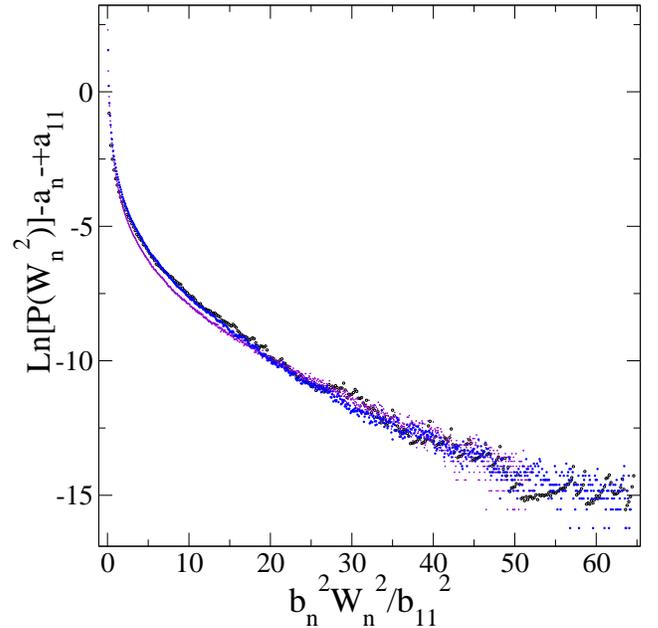}
\caption{Full collapse of the PDF tails of the 11th, 
15th and 18th shells. Note that in the core region the data
does not collapse.}
\label{fig:collapse}     
\end{figure}

To collapse the tails together we need to choose a reference shell
$n_{\rm r}$ ; we show the results for $n_{\rm r}=11$.  Replotting
$\ln[P_n(U_n^2)]-a_n +a_{11}$ as a function of $b_n^2 U_n/b_{11}^2$
one collapses the tails of all the PDF's on the tail of PDF for
$n_{\rm r}=11$.  This is shown in Fig.~\ref{fig:collapse}.

The theoretical predictions~(\ref{eq:U2}, \ref{pred2-PDF}) are
\begin{eqnarray}
  \label{eq:par-scale}
a_n-a_{11}&=&(n-11)(\alpha+\beta-z)\,\ln \lambda\,,  \\ \nonumber
2 \ln (b_n/b_{11})&=& (n-11)\, \alpha\ln \lambda\ .
\end{eqnarray}
According to Eqs.~(\ref{eq:U2}) and the relation~$y+z=1$ one computes
$\alpha+\beta-z=2-3z+\beta$. We plot now the measured (by the best
fits) values of $(b_n-b_{11})/\ln \lambda$ {\sl vs} $(n-1)$. Finding
best linear fits to the resulting plots we compute $\alpha= -0.25 \pm
0.03$. Noticing the independently measured values of $y=0.24\pm 0.01$,
$\zeta_2=0.72\pm 0.01$ we see that our value of $\alpha$ is in
excellent agreement with Eq.~(\ref{eq:U2}); the latter predicts
$\alpha=2y -\zeta_2\approx 0.24 $.
We want next to find the value of $\beta$ from the first of
Eqs.(\ref{eq:par-scale}). Unfortunately the values of $a_n$ are not
computable with the same accuracy as those of $b_n$.  The reason for
this is that the fit formulae picks up the values of the intercepts of
Eq.(\ref{eq:fits}) with much worse precision than the
slopes. Accordingly the plot $(a_n-a_{11})/\ln \lambda$ {\sl vs}
$(n-11)$ is much more scattered than the corresponding plot for the
slopes, and we can only offer a rough estimate of the expected values
of $\beta$, $0.2\le \beta\le 0.6$.

This rough estimate is not satisfactory, and therefore we attempt now
to find a sharper result for $\beta$ using Eq.(\ref{eq:zp}).  In paper
\cite{sabra} we measured the values of $\zeta_p$ for
$p=1,\,2,\,3,\dots 7$. We recognize that these values of $p$ are not
large enough to determine the asymptotic slope of
$\zeta_p$. Nevertheless for a semi-quantitative analysis we can use a
reasonable fit formula for the $\zeta_p$-dependence, for example:
\begin{equation}
  \label{eq:fitz}
  \zeta_p=\frac{p}{3}- \frac{\delta\, p\, (p-3)}{1+\gamma p}\ .
\end{equation}
With this we find the ``best'' values of $\delta$ and $\gamma$ that
agree with the measured values of $\zeta_p$: $\delta\approx 0.092$,
$\gamma\approx 0.725$.  With these values Eq.~(\ref{eq:fitz}) predicts
for $p\to \infty$
\begin{equation}
  \label{eq:zeta-limit}
  \zeta_p \approx  0.56 + 0.21\, p\ .
\end{equation}
According to the prediction~(\ref{eq:zp}) the slope of this dependence
is $y$.  The value of $y$ found above from the inter-shell collapse of
the separated peaks is $y=0.24\pm 0.01$, being in agreement with the
value of $y$ found from the collapse of PDF tails. The value
$y=0.24\pm 0.01$ differs a bit from the slope in
Eq.~(\ref{eq:zeta-limit}). Nevertheless in light of the inaccuracy of
the measured values $\zeta_p$ for large $p$ (originating mostly from
the finite extent of the inertial interval), one cannot trust the last
digits in the numbers of Eq.(\ref{eq:zeta-limit}). We thus consider
the agreement between the estimated values of $y$ more than
acceptable.

Thus we will use the intercept in Eq.~(\ref{eq:zeta-limit}) to
estimate $\beta$. Considering Eq.~(\ref{eq:zp}) the free term
in~(\ref{eq:zeta-limit}) has to be $z-\beta$. With $z\approx 0.74$ we
compute $\beta\approx 0.18$ which is at the borderline of the expected
region [0.2,0.6] found above from collapsing the PDF tails.  Taking
then a value of $\beta\approx 0.2$ allows us to evaluate the number of
peaks $N_n$ in $n$ shell when there were $N_{n-1}$ peaks in the
previous one:
\begin{equation}
  \label{eq:estimate}
  N_n/N_{n-1}=\lambda^\beta \approx 1.15\,, \quad \mbox{for}\
  \lambda=2\,,\ \beta=0.2 \ .
\end{equation}
The conclusion is that peak splitting leads (for $\lambda=2$ and the
chosen value of $a,b,c$) to a 15\% increase of $N_n$ from shell to
shell.

A cursory look at Fig.6 may leave the impression that this is
an underestimate. After all, from one peak in shell 15
the cascade forms four or five peaks in shell 20. A rate of
increase of 15\% would result in a factor of 2, not 5. But
we must rememeber that we talk about peaks {\em of a given
amplitude}, and the peak splitting results in peaks of varying
amplitudes. The counting of peaks of comparable amplitudes 
is more subtle, and the predicted rate of 15\% increase should
be interpreted in the statistical sense, taking many realizations
into account.

\section{Self-similar solutions of the Sabra shell model}
\label{s:theory}
In this section we rationalize the scale-invariant form~(\ref{scform})
on the basis of the equations of motion of the Sabra
model~(\ref{sabra}). The exponent $y$ and the times $t_n$ which appear
in Eq.~( \ref{scform}) are chosen according to
\begin{equation}
\label{times}
y=1-z\,, \quad t_n - t_{n-1}=A\lambda ^{-zn}\,,
\end{equation}
with an arbitrary positive parameter $A$; (note that in \cite{95DG}
there was a salient choice of $A=0$).  These choices are not specific
for the Sabra model; in Refs.~\cite{88Nak,95DG} identical choices were
taken the the Obukhov -- Novikov (ON) and the Gledzer -- Okhitani --
Yamada (GOY) models.  The fist relation follows from simple power
counting, since the RHS of the equation of motion for $n$th shell is
proportional to $\lambda^n$. Indeed, we saw that this scaling relation
is in good agreement with our numerical observations.  The second
choice of~(\ref{times}) reflects the fact the time delay between the
appearance of the peaks in consecutive $n$ shells falls off
geometrically with $n$, and see Fig.~\ref{fig4} as an
example. Nevertheless we want to show directly that these choices are
supported by the equations of motion.

In doing so we follow Ref.~\cite{88Nak}. Substituting~(\ref{scform})
and~(\ref{times}) in~(\ref{sabra}) we find the equation of motion of
the scaling function $f(\tau)$ which is valid in the inertial
interval:
\begin{eqnarray}\label{eq:univ}
\frac{d\, f(\tau)}{d\tau}&=& \rm{St}(\tau)\ , \\ \label{st} 
\rm{St}(\tau)
&\equiv& \lambda ^{3z-2}f^*[\lambda^{z}(\tau-\tau_0)+\tau_0]
f[\lambda^{2 z}(\tau-\tau_0)+\tau_0]\\ \nonumber  &+&\,\,c\lambda
^{2-3z}f [\lambda^{-z }(\tau-\tau_0)+\tau_0] f[\lambda^{-2
z}(\tau-\tau_0)+\tau_0]\\ \nonumber
 &-&(a+c)f^*[\lambda^{-z}(\tau-\tau_0)+\tau_0] f
\,[\lambda^{z}(\tau-\tau_0)+\tau_0]\ .
\end{eqnarray}
To get this equation we changed the time variable from $t$ to $\tau
_n\equiv \lambda ^{nz}(t-t_n)$, and used the same $\tau_n$ in all the
shells involved in (\ref{sabra}), and finally denoted $\tau_n \equiv
\tau $. The characteristic time $\tau_0$ is obtained from computing
the sum of all time increments $\sum_{m=n}^{\infty}(t_{m+1}-t_{m})$,
and noting that it converges to $t_0=\lambda ^{-nz}\tau_0$, where
\begin{equation}
\label{tau0}
 t_0=\lambda ^{-nz}\tau_0\,, \quad \tau_0 \equiv A/(\lambda ^z -1)\ .
\end{equation}
The meaning of $t_0$ is the time needed for a pulse to propagate from
the $n$th shell all the way to infinitely high shells. The
characteristic time $\tau_0$ allows one to convert all the arguments
of the functions $f$ involved in (\ref{st}) to a universal form
$[\lambda^{\alpha} (\tau-\tau_0)+\tau_0]$.

It was shown in Ref.~\cite{88Nak} that the
Eqs.~(\ref{eq:univ},~\ref{st}) can be considered as a nonlinear
eigenvalue problem. They have trivial solutions $f(\tau)=0$, but they
may have nonzero solutions for particular values of $z$ and $A$. For
example, the nonzero solution $f(\tau)=\ $const. requires
$z=2/3$. Nevertheless the constant solution fails to fulfill the
requirement that $\lim_{\tau\to \pm \infty}f(\tau)=0$. We expect that
a nontrivial solution that satisfied the boundary conditions will
force $z$ into the observed value which lies between 2/3 to 1. The
actual calculations that demonstrate this are outside the scope of
this paper.  We just reiterate our numerical finding that $z\approx
0.75$ for the particular set of  parameters $a$, $b$, $c$ and $\lambda$
that were employed in this study.

\vskip -0.6cm \begin{figure}
\epsfxsize=8.4truecm \epsfbox{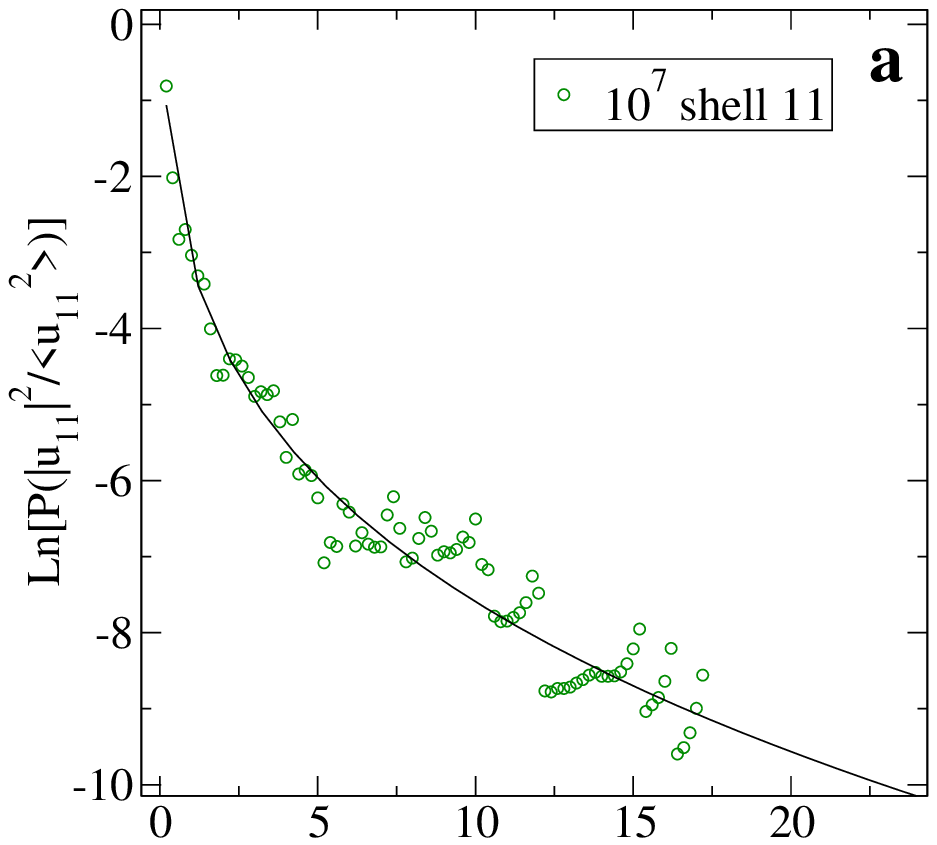}\vskip -0.3cm
\epsfxsize=8.4truecm \epsfbox{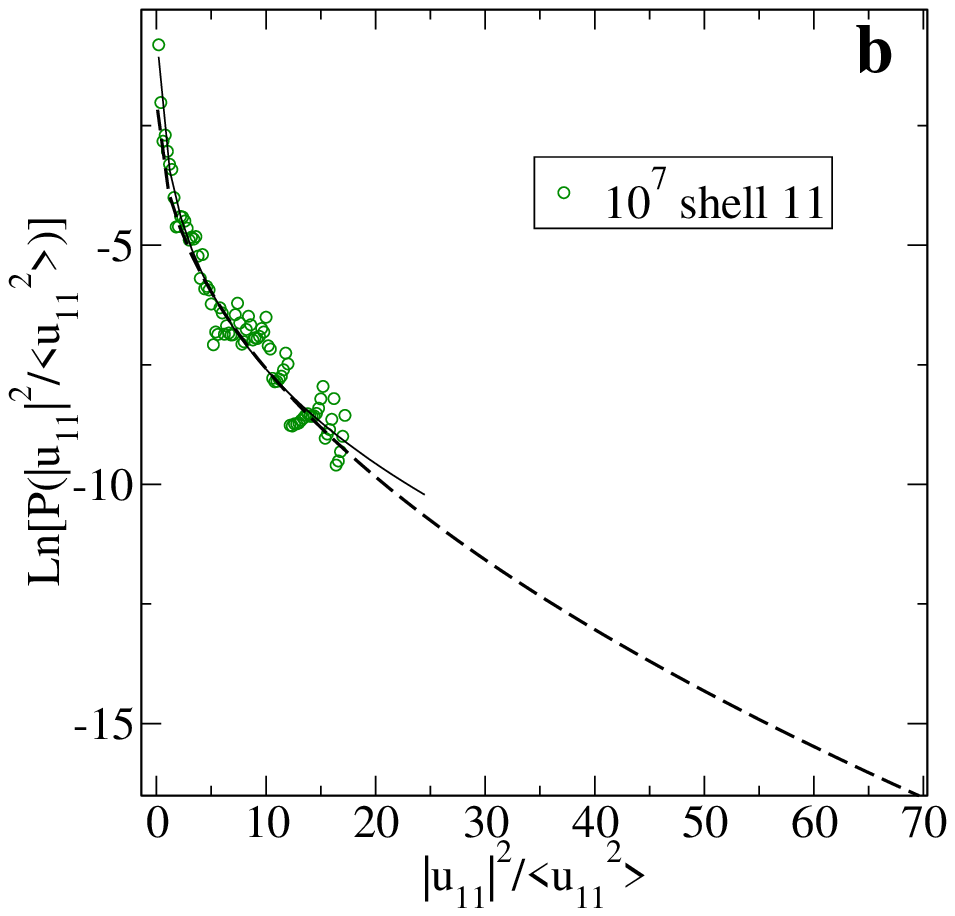}\vskip -0.3cm 
\caption{Panel a: data and analytic fit for the PDF of the 
11th shell in a short time horizon of $10^7$ time steps. Note that
here we presented all the events, including four isolated
events that give rise to the upswingings strings of data points
with amplitudes larger than 7. Panel b: same
as in panel a together with the tail (dashed line) predicted from the
tail of the 18th shell in the same short time horizon. }
\label{predict}
\end{figure}
\begin{figure}
\epsfxsize=8.6truecm \epsfbox{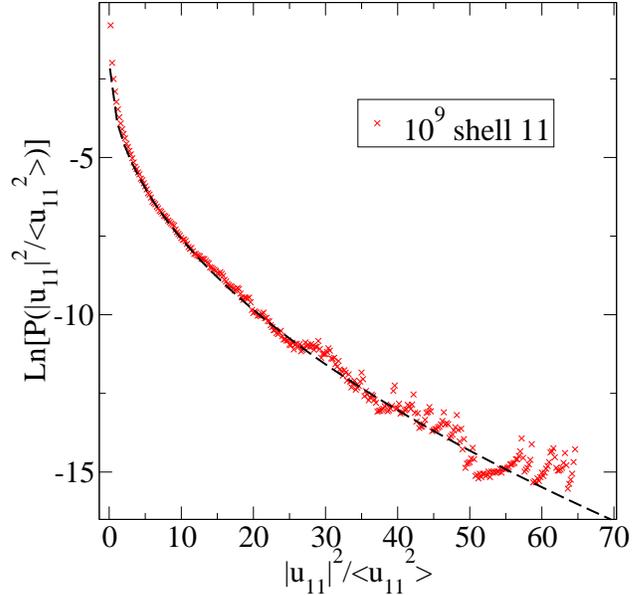}
\caption{
Test of the predicted PDF for the 11th shell using data from a hundred
fold longer time horizon of $10^9$ time steps.}
\label{predictc}
\end{figure}
\section{Predicting tails of PDF's from data measured in short
 time horizons} \vskip -0.3cm In this Section we demonstrate that the
 analysis presented above can be used to predict the tails of the
 PDF's of large scale phenomena (relatively low values of $n$) using
 only data measured in the short time horizon. We focus on the example
 shown in Fig.~\ref{fig2}, i.e. $n=11$ with $10^7$ times steps.

 We first fit the PDF shown in Fig.~\ref{fig2}, 
using a fit formula which is inspired by Eq.~(\ref{eq:fits}):
\begin{equation}
\ln[P_n(u_n^2)]=\tilde a_n+\tilde b_n u_n^{\tilde c_n} \ ,
\label{fit11}
\end{equation}
\label{sec:conclusion}
and found $\tilde a_{11}\approx 1.34$, $\tilde b_{11} \approx -4.64$.,
$\tilde c_{11} \approx 0.28$. The data and the best fit are shown in
Fig.\ref{predict} panel a.

Next we want to continue the PDF of $n=11$ into event values that are
too rare in the short time horizon. To this aim we measured, in the
same time window of $10^7$ time steps, the tail of the PDF of the 18th
shell. In doing so we use the fact that the small scale events have a
much shorter turn over time, and the ``short" time horizon is
sufficiently long to provide a good estimate of the tail.  We fitted
the tail with Eq.~(\ref{eq:fits}) and found $a_{18}\approx -5.3$,
$b_{18}\approx -0.94$. From this value and (Eq.~\ref{eq:par-scale}) we
can predict $b_{11}$. We employ the value $\alpha \approx 0.24$ which
is taken from Eq.(\ref{eq:U2}) with the known value of $y$ (from the
intershell collapse) and of $\zeta_2$. The resulting  prediction is
$b_{11}\approx -1.72$.

Rather than attempting to also predict $a_{11}$ in Eq.~(\ref{eq:fits})
(knowing the inaccuracies of intercepts) we glued the tail with the
predicted value of $b_{11}$ to the core PDF function (\ref{fit11}) by
finding the unique point of continuity with same first derivative.
The way that the predicted tail hangs onto the PDF is shown in
Fig.~\ref{predict} panel b.

To test the quality of the prediction we ran now the simulation for a
time horizon that is a hundred times longer (i.e $10^9$ time
steps). Such a run can resolve the events that belong to the tail, and
indeed the comparison is surprisingly good, as seen in
Fig.~\ref{predictc}.

\section{Summary}

The main aims of this paper are twofold: on the one hand we
aimed at understanding the detailed dynamical scaling properties
of the largest events in our system. On the other hand we wanted
to employ these properties to {\em predict} the probability
of these events even in situations in which they are very rare.

The first aim was achieved by focusing on the largest events,
following their cascade down the the scales (or up the shells), and
learning how to collapse them on each other by rescaling their
amplitudes and their time arguments. This exercise culminated in
Eq.(\ref{scform}) which represents the largest events $u_n(t)$ in
terms of a ``universal" function $f(\tau)$ where $\tau$ is a properly
rescaled time difference from the peak time of the event. This
dynamical scaling form is characterized by two exponents, a ``static"
one denoted $y$ and a ``dynamic" one denoted $z$. We argued
theoretically for a scaling relation $z+y=1$, and determined the
values of the these exponents on the basis of the analysis of isolated
events in {\em short} time horizons.

The second aim was accomplished by developing a scaling theory for the
tails of the PDF's in different shells. We have learned how to translate
information from the tail of a PDF in a high shell to the tail of a
PDF of a low shell. In doing so we made use of the fact the high
shells (small length scales) have much shorter characteristic times
scales. Thus even short time horizons are sufficient to accumulate
{\em reliable} statistics on the tails of the PDF's of high
shells. Having a theory to translate the information to low shells in
which the tails are extremely sparse (or even totally absent), we
could overcome the meager statistics. We could present predicted tails
that were populated only in time horizons that were a hundred fold
longer than those in which the analysis was performed.

We demonstrated the existence of scaling properties of the extreme
events that are in distinction from the bulk of the fluctuations that
make the core of the PDF. In this sense the extreme events are
outliers.  We cannot, on the basis of the present work, claim that
this approach has a general applicability to a large class of physical
systems in which extreme events are important. We certainly made a
crucial and explicit use of the scale invariance of the underlying
equation of motion. This scale invariance translates here to an
intimate connection between extreme events appearing on one length
scale at one time to extreme events appearing on smaller length scales
at later (and predictable) times (cf. Fig.6). We are pretty confident
that similar ideas can (and should) be implemented to fluid
turbulence; whether or not such techniques will be applicable to
broader issues like geophysical phenomena or financial markets is a
question that we pose to the community at large.

\acknowledgments Our interest in the issue was ignited to a large
extent by the meeting on extreme events organized by Anne and Didier
Sornette in Villefranche-sur-mer, summer 2000. We thank Didier
Sornette for his comments on the manuscript, and for clarifying the
notion of ``outliers". This work has been supported in part by the
European Commission under the TMR program, the Israel Science
Foundation, The German Israeli Foundation and the Naftali and Anna
Backenroth-Bronicki Fund for Research in Chaos and Complexity.
 \end{multicols}
\end{document}